# Residual Stress-Driven Non-Euclidean Morphing in Origami Structures

Zihe Liang[1†], Sibo Chai[2,3†], Qinyun Ding[1], Kai Xiao[1], Ke Liu[4], Jiayao Ma[2,3*], Jaehyung Ju[1*/]

[1]UM-SJTU Joint Institute, Shanghai Jiao Tong University, 800 Dongchuan Road, Shanghai 200240, China

[2]Key Laboratory of Mechanism and Equipment Design of Ministry of Education, Tianjin University, Tianjin 300350, China

[3]School of Mechanical Engineering, Tianjin University, Tianjin 300350, China

[4]Department of Advanced Manufacturing and Robotics, College of Engineering, Peking University, Beijing 100871, China

**Abstract:**

Non-Euclidean surfaces are ubiquitous in numerous engineering fields, such as automotive, aerospace, and biomedical engineering domains. Morphing origami has numerous potential engineering applications, including soft robots, mechanical metamaterials, antennas, aerospace structures, and biomedical devices, owing to its intrinsic morphing features from two-dimensional (2D) planes to three-dimensional (3D) surfaces. However, the current one-dimensional (1D) hinge deformation-driven transformation of foldable origami with rigid or slightly deformable panels cannot achieve a 3D complex and large curvilinear morphing. Moreover, most active origami structures use thin hinges with soft materials on their creases, thus resulting in a lower load capability. This study proposes a novel origami morphing method that demonstrates large free-form surface morphing, e.g., Euclidean to non-Euclidean surface morphing with shape-locking. We embedded tensorial anisotropic stress in origami panels during the extrusion-based 3D printing of shape memory polymers. The extrusion-based 3D printing of isotropic shape memory polymers can produce tensorial anisotropic stress in origami panels during fabrication, which can realize large non-Euclidean surface morphing with multiple deformation modes. The connecting topology of the origami unit cells influences the global morphing behavior owing to the interaction of the deformation of adjacent panels. Non-Euclidean morphing integrated with four-dimensional (4D) printing can provide multimodal shape locking at material and structural levels. The non-Euclidean surface morphing caused by tensorial residual stress in the panel during 3D printing expands the design space of origami and kirigami structures.

# I. Introduction

Morphing shapes from a two-dimensional (2D) plane to a three-dimensional (3D) space can significantly influence future manufacturing, such as in Industry 4.0, which requires customized mass production and convenient logistics for storage and transportation[1,2]. Moreover, morphing complex geometries from a 2D plane to a 3D curved non-Euclidean surface can reduce manufacturing costs by directly embedding the design into the manufacturing process. The morphing involved in manufacturing does not require additional labor to assemble 3D shapes owing to the single-step fabrication followed by morphing[3–5]. Morphing does not require the use of supporting materials for the 3D printing of overhang features. This reduces fabrication energy requirements by increasing the manufacturing speed owing to the reduction of fabrication volume without use of supporting materials[6,7]. The physical properties and functionality of the morphed shapes can be tuned as a byproduct. The morphing strategy can be more useful in micro- and nano-fabrication, where the direct fabrication of 3D complex geometries is challenging. However, 2D plane-based fabrication is more conventional, e.g., electron beam nanolithography[8–10] and atomic layer deposition[11,12].

Owing to the intrinsic features of the origami (or kirigami), folding of a 2D plane shape into a 3D free-formed one, morphing is especially beneficial for fabricating origami structures. The high-level foldability of origami structures enables the production of self-foldable structures and active metamaterials. However, current origami structures have a limitation of morphing into non-Euclidean surfaces that are ubiquitous in the automotive, aerospace, and ocean engineering fields[13]. Despite the existing applications of origami in robotics[14,15], deployable structures[16,17], and biomedical devices[18,19], the transformed configurations of origami structures have retained flat plates, thus globally generating a discretized geometry due to the use of simple one-dimensional (1D) deformation mechanisms on the hinges such as bending/folding[20,21] or twisting[22].

Four-dimensional (4D) printing is a common method for demonstrating the morphing of simple origami structures by integrating functional materials and additive manufacturing[21,23]. Embedding entropy, e.g., internal stress, in the 3D printing of smart materials, is a typical strategy in 4D printing[24–26]. However, most 4D printing techniques are limited with respect to the synthesis of strong, morphable, and 3D curvilinear structures. Most membrane-based 4D printing methods with a heterogeneous distribution of scalar entities with physical or chemical potential[27,28] cannot be scaled up to structural applications owing to the extremely low stiffness of the base materials ($E < \sim 0.5 MPa$) and thin membrane dimension[23,27,29]. Without embedding spatial anisotropy with reinforcement and shape locking during fabrication, 3D curvilinear structures cannot be realized using scalar field gradient-based 4D printing[27,28,30,31].

To address the dual challenges of active origami and 4D printing, reforming the conventional transformation strategy of origami from hinges to panels is necessary; this can be achieved by programming using high-level physical intelligence during additive manufacturing. In this study, tensorial anisotropic residual stress was embedded into a thick panel during the 3D printing of a shape memory polymer to morph a flat surface into those with nonzero Gaussian curvatures, while exhibiting dual shape-locking modes at the material and structural levels. This work demonstrates that tensorial anisotropic stress embedded in a flat plate[31–33] during 3D printing can generate 3D curvilinear self-foldable origami structures with multiple deformation modes and connection effects.

## II. 4D printing of non-Euclidean surfaces

Most reconfigurable structures, including origami, use hinge deformation with simple folding/bending modes[20,21,34] to create discrete and noncurvilinear geometries. The hinge deformation-driven self-folding origami harnesses a vector field controlled by the temperature or moisture gradient. Unlike the hinge-driven folding shown in Figure 1a, embedding a tensor field into panels can produce complex morphing shapes while transforming the flat panels into a non-Euclidean surface, as demonstrated by the manufacture of the Yoshimura origami patterns shown in Figure 1a.

Non-Euclidean morphing can be implemented using extrusion-based 3D printing such as direct ink writing (DIW)[35–37] or fused deposition modeling (FDM)[23,24]. Figure 1b illustrates the principle of morphing non-Euclidean surfaces by the extrusion of a single material. The extruded filaments of shape memory polymers (SMPs) are pre-stressed in the longitudinal direction with Poisson's effect in the lateral direction during the printing process above the melting temperature. This is analogous to the typical mechanical training of SMPs above their glass transition temperature ($T_g$). The printed flat panel retains its tensorial internal stress and is fixed at room temperature. Reheating above the $T_g$ of the SMP causes the filaments to deform, thus transforming them to a low-energy state by releasing energy while exhibiting a shape-memory effect. Combining this principle with multilayer printing in different printing directions can introduce more complex tensorial residual stress during printing, thus producing a morphed structure with complex surface shapes, such as doubly curved surfaces with negative Gaussian curvatures. Notably, the pre-stressed filaments layered with $[\theta_{top}, \theta_{botton}] = [45°, -45°]$ produce a saddle-shaped surface, as illustrated in Figure 1c. Moreover, the morphed structure exhibits two shape-locking modes. The first is the locking mode of the material with the shape-fixity effect of the SMP when the temperature decreases after the initial thermal deployment above $T_g$. The second is structural locking due to mechanical instability, which is discussed in Section V. After the initial thermal deployment, the shape of the bistable curved tube changes to a different low-energy state through mechanical loading, as shown in Figure 1d. The bistable structure produces two different lockable modes that are longitudinally and laterally transformed by the connection topology of the panels.

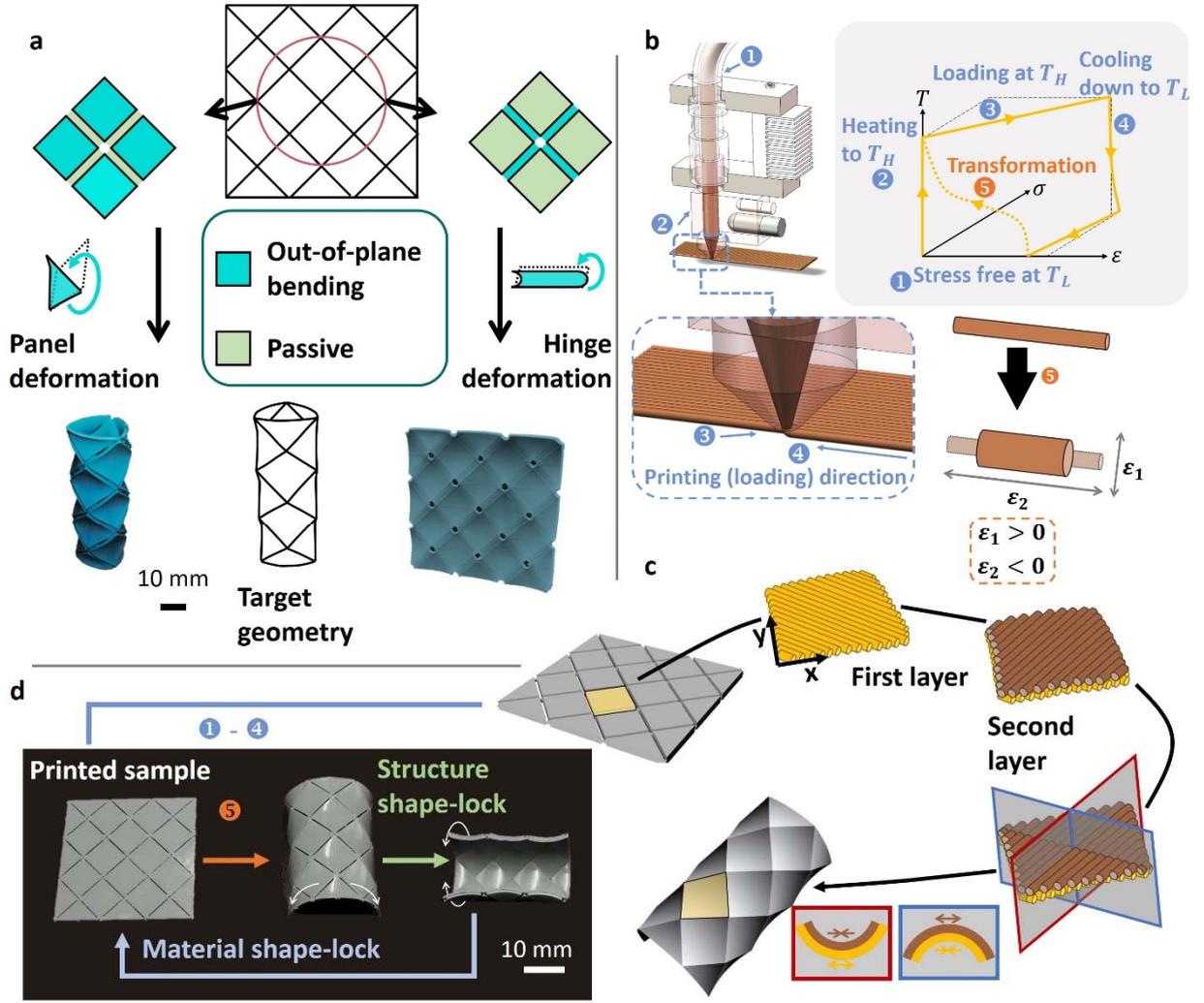

*Figure 1. **Panel deformation-based self-foldable origami manufacturing**, a. Comparison of two origami manufacturing methods for the fabrication of a Yoshimura pattern: the conventional soft hinge-driven transformation has a limitation in transforming complex structures. Moreover, the panel deformation can produce more complex transformations such as a non-Euclidean surface and large deformation. **b.** 4D printing of a single material to produce anisotropy and the shape memory effect. **c.** Programming with tensorial stress on the panel to morph into a non-Euclidean structure, e.g., a Yoshimura pattern, with doubly curved surfaces, **d.** Experimental demonstration of self-foldable origami structures with multimodal shape locking.*

## III. Thermomechanical modeling of panel deformation

To demonstrate panel deformation programming using direct 4D printing, we developed a theoretical model for the thermomechanical shape morphing of a flat panel. Subsequently, we implemented a mathematical relationship to describe the connectivity between the individual and adjacent plates during shape morphing.

We adopted the classical laminated plate theory (**CLPT**), where we obtained the in-plane strain ($\boldsymbol{\varepsilon}^0$) and out-of-plane curvature ($\boldsymbol{\kappa}$) by computing the effect of the external resultant loadings, force (**N**), moment (**M**), thermal resultant loadings, thermal force ($\mathbf{N^T}$), and thermal moment ($\mathbf{M^T}$) on the printed bilayer laminated composites. The material properties and layer (or ply) orientations of the bilayer composite contribute to **A**, **B**, and **D**, which represent the extensional, bending-extensional, and bending stiffness tensors, respectively. The superscript '0' of the strain indicates the midplane of the plate. The thermomechanical constitutive relationship is expressed as follows[38]:

$$\begin{Bmatrix} \mathbf{N} \\ \mathbf{M} \end{Bmatrix} = \begin{bmatrix} \mathbf{A} & \mathbf{B} \\ \mathbf{B} & \mathbf{D} \end{bmatrix} \begin{Bmatrix} \boldsymbol{\varepsilon}^0 \\ \boldsymbol{\kappa} \end{Bmatrix} - \begin{Bmatrix} \mathbf{N}^T \\ \mathbf{M}^T \end{Bmatrix}. \tag{1}$$

Notably, the three stiffness tensors ($\mathbf{A}$, $\mathbf{B}$, and $\mathbf{D}$) are determined by the anisotropic material modulus and Poisson's ratio of the SMP material, in addition to the printing pattern of each layer. The resultant thermal loadings are determined from the longitudinal shrinking strain of the SMP fibers when considering Poisson's effect. We adopted polylactic acid (**PLA**) as an SMP material, with its longitudinal shrinkage values dependent on the exposed time and actuation temperature; as adopted from prior research[24,39,40] for a nozzle speed of $60\ mm/s$, nozzle temperature of $210\ °C$, and layer height of $0.1\ mm$. The expressions for each term are expressed by Equation (S9) in Section S1 of the Supplementary Information (SI), and the material properties in Section S3 of the SI.

The Green–Lagrange strain tensor $\boldsymbol{\varepsilon}\ (=\boldsymbol{\varepsilon}^0 + z\boldsymbol{\kappa})$ is a combined function of $\boldsymbol{\varepsilon}^0$ and $\boldsymbol{\kappa}$, where $z$ is the distance from the midplane of the composite in the thickness direction. The components of the displacement vector $\mathbf{u}$ can be expressed as $u_i = u_i{}^0 - z\frac{\partial u_3{}^0}{\partial X_i}$ for $i = 1,2$ and $u_3 = u_3{}^0$, where $X_i$ is the basis vector components and $u_i{}^0$ is the midplane displacements. From the strain–displacement relationship, we can obtain $\mathbf{u}$ to predict the deformed configurations of a single plate:

$$\boldsymbol{\varepsilon} = \frac{1}{2}[(\boldsymbol{\nabla}_x \mathbf{u})^T + \boldsymbol{\nabla}_x \mathbf{u} + (\boldsymbol{\nabla}_x \mathbf{u})^T \boldsymbol{\nabla}_x \mathbf{u}]. \tag{2}$$

To solve Equation (2), we imposed boundary conditions $\mathbf{u} = const$ on $\varGamma_u$, where $\varGamma_u$ is the surface boundary. The explicit expression of the boundary condition of the laminated segments varies with the printing direction of the layer, and the boundary condition of the laminates is presented in Section S1 of the SI. Figure 2a illustrates the thermomechanical shape morphing of a single $[45°, -45°]$ laminated rhombus plate, which was initially flat and then deformed to a doubly curved surface in a non-Euclidean space. In particular, $a$, $b$, and $d$ denote the geometric parameters in the undeformed configuration; and $\bar{a}$ is the length of a deformed plate, whereas $\bar{b}$ and $\bar{d}$ are the diagonal lengths along the lateral and longitudinal directions, respectively.

## IV. Connecting topology effect on morphing

### *Assembly with connection of vertices*

Considering the continuity of individual segments, the global deformation of the assembled panels can be estimated when the deformed configurations of the panel are known. For example, a rhombus panel with a negative Gaussian curvature can be 4D printed, as shown in Figure 2a. The in-plane tessellation of a $[45°, -45°]$ laminated panel with a connection at the vertices can generate a cylindrical shape such as a Yoshimura pattern[41] with a global zero-Gaussian curvature, as shown in Figure 2b. The deformed configuration of an assembly of rhombus panels can be expressed as follows:

$$\tan\frac{\varphi}{4} = \frac{\sqrt{4\bar{a}^2 - \bar{b}^2 - \bar{d}^2}}{\bar{b}}. \tag{3}$$

Notably, $\varphi$ characterizes the circumferential deformation of a single rhombus panel. The deformation magnitude of the assembly can be expressed as a global curvature $\kappa_g$, which is defined as $\kappa_g = 1/r_g$, where $r_g$ denotes the radius of the global curvature generated by the interaction of individual panels and hinges. Equation (3) reveals that the global curvature is a function of the rhombus angle $\alpha$. The explicit definition of $r_g$ is expressed in Section S1 of SI.

Figure 2c presents the geometric parameters of the assembly, including the hinge length $l_c$, hinge width $d_c$, panel thickness $t_p$, and hinge thickness $t_c$. The deformed shape in Figure 2b was generated with $t_p = 2\,mm$, $t_c = 0.4\,mm$, $l_c/d = 0.2$, and $d_c/d = 0.05$ for $d = 16\,mm$. The local deformation of the individual panels interacted with the hinges to form a globally deformed shape. Figure 2e presents the deformation of the hinges along the circumferential direction of the cylinder when the panels were connected at the vertices. Notably, a panel with $\alpha = 90°$ produces a doubly curved surface with two principal directions: circumferential and longitudinal. Global deformation occurred along the circumferential direction, which requires less energy than longitudinal deformation. The hinges generated a convex shape along the circumferential direction to align with the convex principal curvature of the panels, as shown in Figure 2e. However, the thermomechanical strain energy did not overcome the energy barrier in the longitudinal direction, where the concave principal curvature shape of the panel did not influence the global shape. Moreover, only the hinges deformed to match the continuity of Elastica[42].

The thickness ratio $R_{lt}(= t_1/t_p)$ of the panel layups can influence the local curvature $\kappa$ of panels and $\kappa_g$. Figure 2f presents $\kappa_g$ values of assemblies with $\alpha = 90°$ for varying $R_{lt}$ values, presenting a comparison of the values for analytical models in Section S1 of the SI, finite element (FE) simulations, and experiments with 3D scanned data. Figure 2g presents a design map of $\kappa_g$ for varying $\alpha$ and $R_{lt}$ values. Notably, the assembly with $R_{lt} = 0.5$ and $\alpha = 90°$ exhibited a bistable property, as shown in Figure 1d and Supplementary Video S3; this is further discussed in Section V.

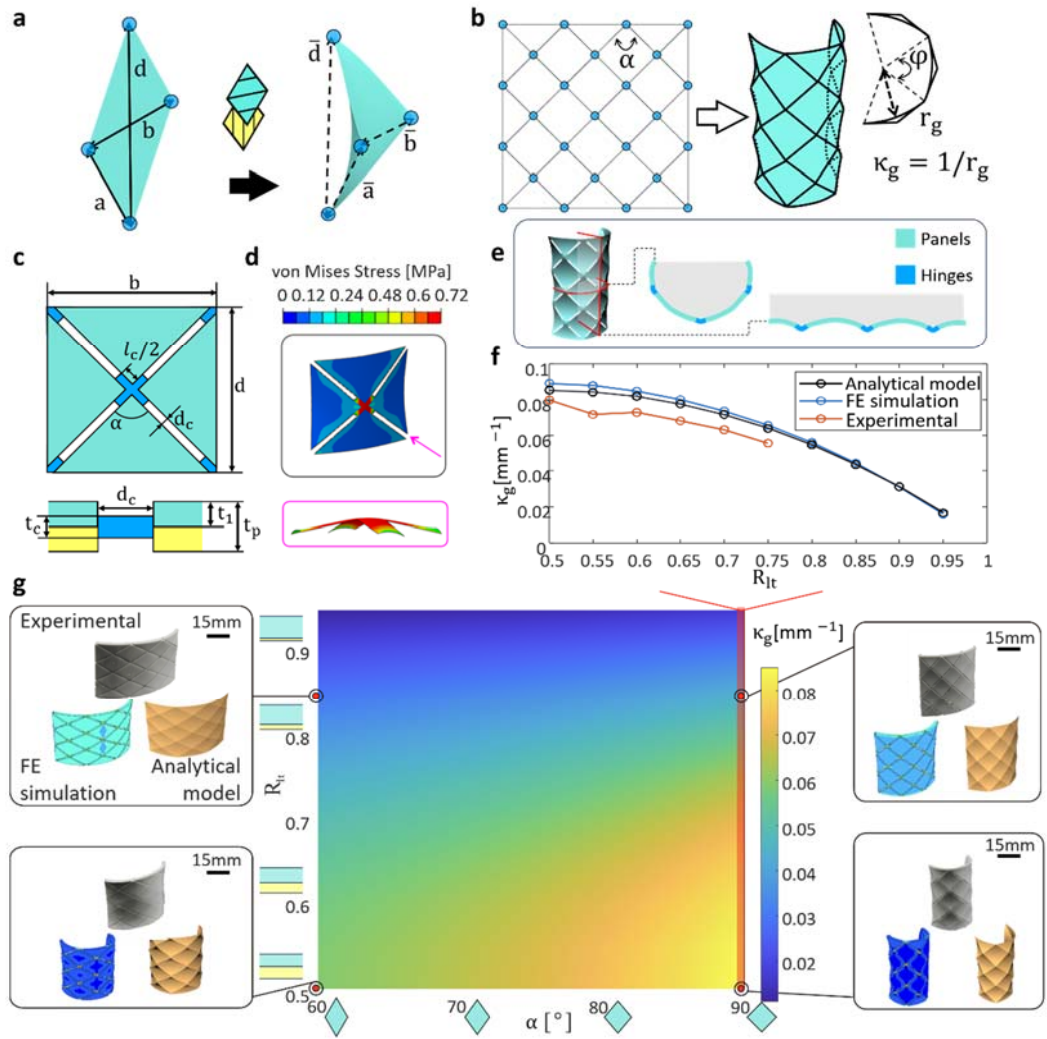

Figure 2. **Thermomechanical deformation of a vertex-connected pattern. a.** Deformation of a single rhombus unit. **b.** Deformation of a 3 × 3 vertex-connected pattern. **c.** Geometric parameters of the vertex-connected pattern. **d.** Local crease deformation of the vertex-connected pattern with parameters ($\alpha = 90°$ and $R_{lt} = 0.5$). **e.** Deformed configuration of a 3 × 3 vertex-connected pattern with crease design, and its cross-section profiles **f.** Comparison of the global curvature $\kappa_g$ for varying thickness ratios $R_{lt}$ in the analytical model, FE simulation, and experiment. **g.** Design map for $\kappa_g$ obtained by the analytical model for varying $\alpha$ and $R_{lt}$ values.

## Assembly with edge connection

In addition to the connection of the vertices, we investigated the topology effect of an assembly on the global deformation by changing the link location from the vertices to the mid-edges, as shown in Figure 3a. Figure 3b presents a comparison of the $\kappa_g$ values of assemblies with 3 × 3 units for the vertex-connected and mid-edge connected conditions with respect to varying $R_{lt}$ and $\alpha$ values, which indicates that a mid-edge connection produces a lower $\kappa_g$ due to the hinge alignment along the off-principal curvature directions of deformed panels. The mid-edge hinges have a torsional deformation mode to satisfy the continuity of the adjacent deformation of panels along the off-diagonal curvature directions, and Figure 3c presents evidence of a lower $\kappa_g$ in Figure 3b.

Similar to the vertices connection, the assembly with $R_{lt} = 0.5$ and $\alpha = 90°$ in the mid-edge connection exhibited a bistable property by the competition of deformation between the longitudinal and circumferential directions with instability. Figure 3d presents a further parametric study of $\kappa_g$ with respect to two varying geometric parameters: the hinge length $l_c$ and hinge thickness $t_c$ (or $\alpha = 90°$ and $R_{lt} = 0.5$). In particular, a greater $l_c$ and smaller $t_c$ produced a high $\kappa_g$, which is related to simple torsional deformation in the hinges.

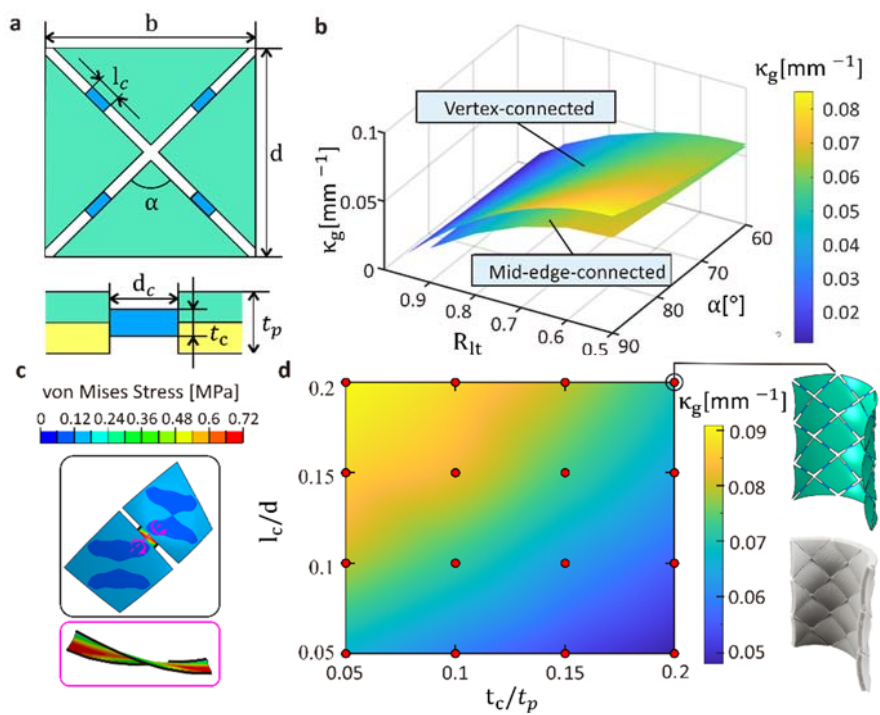

Figure 3. **Thermomechanical deformation of an edge-connected pattern; a.** Geometric parameters of the edge-connected pattern. **b.** Comparison of global curvature for the vertex-connected and the mid-edge-connected patterns for varying $\alpha$ and $R_{lt}$ values. **c.** Local crease deformation of the structure with parameters of $\alpha = 90°, R_{lt} = 0.5, t_c/t_p = 0.2, and\ l_c/d = 0.2$. **d.** Relationship between the normalized crease thickness and crease length with respect to the global bending of the 3 × 3 units, in addition to the deformed configuration and local crease deformation of the structure with parameters of $t_c/t_p = 0.2\ and\ l_c/d = 0.2$.

## V. Bistability of 4D printed origami

A rhombus unit with $R_{lt} = 0.5$ and $\alpha = 90°$, as shown in Figure 2a, thermally morphed into a surface with negative Gaussian curvature. The FE simulations and experiments demonstrated no snapping bistability of the saddle geometry transformed from the rhombus unit, as presented in Section S4 of the SI and Supplementary Video S8. However, when the rhombus unit was connected to adjacent units and thermally transformed, it was bistable. Figure 4a presents a phase map of the bistability of an extended unit comprising four halves of rhombus panels connected to the vertices at the center and four corners. For varying $\alpha$ and $R_{lt}$ values, the phase map obtained from FE simulations exhibited three distinctive regions: bistable, monostable, and no negative stiffness. A greater $\alpha$ and smaller $R_{lt}$ provide a higher probability of the bistability of the extended unit. Figure 4b presents the vertical force–deflection curves of the extended units after thermal transformation. We obtained a reaction displacement for a vertical force at the top center, whereas the four corner nodes were allowed to slide in the plane direction. Figure 4b presents three representative force–deflection behaviors, i.e., bistable $(B_2/B_1 < 0)$, monostable $(0 \leq B_2/B_1 \leq 1)$, and monotonic $(B_2/B_1 > 1)$; where $B_1$ and $B_2$ are the local maximum and minimum loads, respectively. See Supplementary Video S2 for the deformation animation.

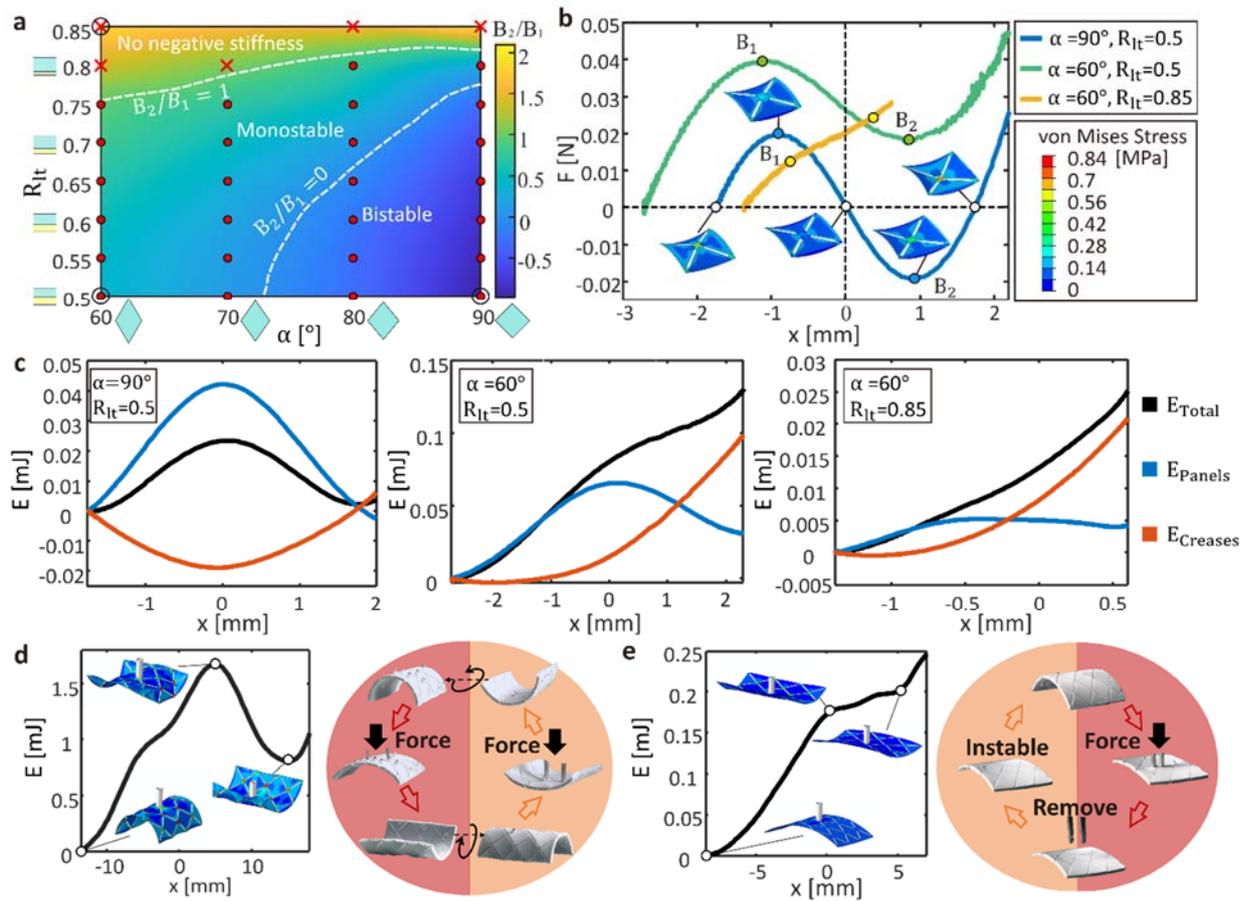

Figure 4. **Bistability of the vertex-connected pattern.** *a. Design map of bistability obtained from FE simulations for varying α and $R_{lt}$ values. b. Representative force–displacement curves of bistable, monostable, and monotonic states. c. Energy composition: total energy, energy by panel deformation, and energy by crease deformation in three representative units. d. Force–displacement curve of a bistable 3 × 3 vertex-connected assembly (α = 90° and $R_{lt}$ = 0.5) and its experimental validation. e. Force–displacement curve of a monostable 3 × 3 vertex-connected pattern (α = 60° and $R_{lt}$ = 0.85) and its experimental validation.*

By decomposing the total energy into those of panels and creases during the snapping process, we found that the total strain energy was in accordance with the crease energy pattern instead of the panel energy pattern, as shown in Figure 4c, which confirms that the connection effect of the rhombus units significantly influences bistability. Figure 4d illustrates the bistable snapping phenomenon of the 3 × 3 assembly with $\alpha = 90°$ and $R_{lt} = 0.5$, whereas Figure 4e demonstrates the monostable snapping for $\alpha = 60°$ and $R_{lt} = 0.85$ (Supplementary Video S3). The bistable and monostable behaviors of the assemblies were in accordance with those of the extended unit cells.

## VI. Multimodal geometric morphing and stiffness

Panel-deformation-driven origami can be extended to multimodal deformations. Figure 5a–5d demonstrate the direct 4D printing of a human mask and the wrapping pattern conceptualized by the thermomechanical composite laminate model integrated with the anisotropically extruded SMP. By encoding five and three deformation modes for the human mask and wrapping pattern, respectively, the printed 2D plates were morphed into 3D complex shapes with individual curvatures after exposure to high temperatures, as shown in Figure 5b and 5d. Notably, the large and multimodal deformations of individual panels were independently controlled, whereas the connection at the vertices adjusted the global shapes.

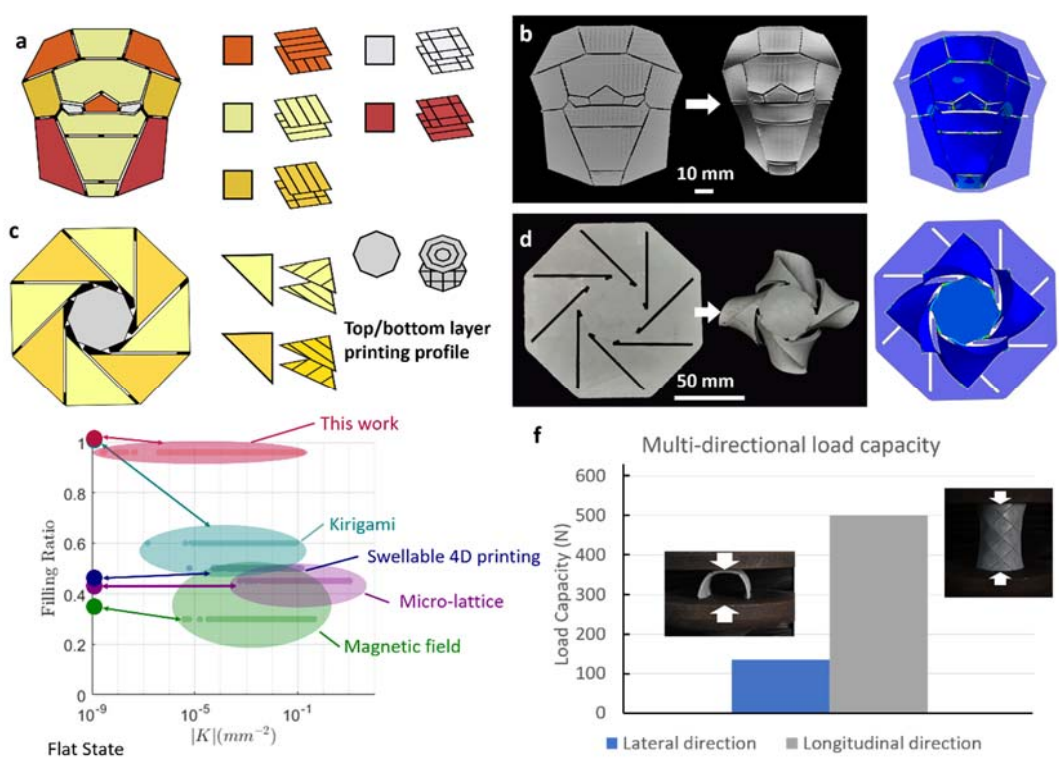

*Figure 5.* **Complex panel-deformation-driven morphing with multimodal deformation and high stiffness;** *a. Design of a human mask with panels that are individually encoded with different laminate layups. b. Experiment and FE simulation of the thermomechanical deformation of the human mask c. Design of a wrapping pattern with panels that are individually encoded with different laminate layups. d. Experiment and FE simulation of the thermomechanical deformation of a wrapping pattern. e. Comparison of the Gaussian curvature ranges for the filling ratios of varying structures with a Kirigami tessellation approach[43], hydrogel 4D printing approach[23], micro-lattice design[3], and Lorentz force-based metasurface[44], where the solid circles on the vertical axis denote the initial flat state. f. Load carrying capacity of panel-deformation-driven origami structures in the lateral and longitudinal directions.*

Implementing large deformations with high stiffness is practically challenging. Low-stiffness materials such as elastomers and hydrogels[27,30] or structures with high porosity, such as kirigami and lattice materials[3,23,43,44], can be used to readily implement large deformations; however, they are weak for structural applications. The panel-deformation-driven morphing, in addition to the shape fixity of the SMP in this study, can increase the filling ratio for high stiffness while exhibiting shape-morphing capacity for complex 3D curvilinear surfaces, as shown in Figure 5e (See Section S8 of SI for details). Kirigami structures are highly compact before morphing, and lose their filling ratios after morphing[43]. However, the origami with panel-deformation-driven morphing did not undergo a significant change in filling ratio after morphing, thus maintaining its structural potential after morphing, as shown in Figure 5e. Figure 5f demonstrates the high load capacity of a Yoshimura patterned structure ($R_{lt} = 0.5$, $\alpha = 90°$) in the longitudinal and lateral directions. Compared with the loading capacity of other structures[3,45], the proposed structure exhibited a higher capacity than the origami and kirigami structures by a factor of approximately 10. See Section S8 of the SI for more details on mechanical testing.

## VII. Discussion

To realize the non-Euclidean surface deployment of the origami structures with large deformation and high stiffness, the proposed method overcomes conventional hinge-based origami folding[21,34,46]. The anisotropic stress field embedded with the shape memory effect of polymers into panels during extrusion-based additive manufacturing can produce a 3D curvilinear deployment of origami structures with varying Gaussian curvatures. The unique method proposed in this study can realize multimodal shape locking at the material and structural levels, which can contribute to numerous potential applications of active origami structures with relatively high stiffnesses, such as robotic metamaterials[46,47], antennas[26], solar arrays[16], morphable automotive structures[27], and biomedical devices[19,48].

Most studies on shape morphing from the 2D plane to 3D space have been conducted with cellular structures[3,26,43] or soft materials[23,27] to realize a larger morphing capacity. These materials cannot be used for scaled-up structural applications owing to their minimal load-carrying capacity. Moreover, although the high foldability of origami (or kirigami) provides a solution for the structural design of 2D–3D shape morphing, current highly discretized origami with flat panels cannot match the morphing capabilities of cellular structures. Notably, the proposed thick panel-deformation-driven active origami can be used to design thick panels with stiffer materials, which contribute to a higher global stiffness of structures while providing a relatively wide range of shape morphing (Figures 5e and 5f). The SMP shape memory and shape fixity effects can facilitate the extensive structural transformation and load-carrying capabilities of origami structures, where there are differences of three orders of magnitude between the transformation and load-carrying modes.

Conventional origami structures with negative Gaussian curvatures are bistable because the traditional folding method requires the instability of the doubly curved geometry from a flat plane[49]. Notably, the panel-deformation-driven active origami structures produce instability-free Gaussian curvatures because the panel deformation mode by 4D printing is bending and not snapping, as shown in Supplementary Video S8. However, the interactions of adjacent instability-free Gaussian curvatures with the hinge topology can produce bistable structures, as shown in Figure 4.

Although the proposed active origami structures can produce large deformations of thick panels with various Gaussian curvatures, the current thin-panel-based model expressed by Equation (1) may be rendered inaccurate for the prediction of the curvature with an increase in the panel thickness. In this study, an inverse design method was not provided for the deformation of an assembly of panels, which will be investigated in future research.

## VIII. Conclusion

In summary, this study introduced a non-Euclidean morphing method of origami, which is integrated with an extrusion-based 4D printing of shape memory polymers, to demonstrate the self-deployment of origami with 3D curvilinear shapes. The programming of tensorial anisotropic residual stress into panels from isotropic shape-memory polymers during extrusion-based additive manufacturing using the thermomechanics of printed laminates can realize non-Euclidean morphing with different deformation modes, including a doubly curved surface. Combining the unique shape fixity of SMP with mechanical bistability can produce multimodal shape locking at the material and structural levels. Potential applications include deployable and tunable automobile, aerospace, and ocean engineering structures designed to reduce fluid friction on smooth surfaces. The proposed approach represents reversible, shape-lockable, and scalable morphing structures, for which conventional reconfigurable structures are yet to be developed.

## IX. Materials and Methods

**Sample fabrication and deployment:** This work uses a Fused Deposition Modeling (FDM) 3D printer (Ultimaker 2+, Ultimaker, The Netherlands) with a commercially available polylactic acid (PLA) filament (PolyLite PLA, Polymaker) to print flat origami samples. We construct 3D models of the origami samples with a 3D CAD design software (Solidworks 2021, Dassault Systèmes), incorporating geometric parameters of origami structures such as the thickness ratio, rhombus angle, and unit connection conditions. We import 3D models into a preprocessing code of our FDM 3D printer (Ultimaker Cura 5.0.0, Ultimaker, The Netherlands), where they are sliced into G-Code files that direct the 3D printing process. We adjust the printing angles in Cura's 'Infill' section to achieve the bilayer fabrication of origami panels. Other essential printing parameters include the nozzle temperature (180 °C), printing-layer height (0.1 mm), and print speed (50 mm/s). After printing, we deploy the samples in a hot water tank where an immersion heater (Anova, USA) maintains and monitors the temperature. The morphed origami structures made of PLA above its glass transition temperature ($T_g \geq 60°C$) lock their shapes when the temperature cools down to room temperature ($T_r \sim 20°C$).

**Measurement of panel deformation:** We validate the analytical and numerical models with experiments. A 3D scan system (Einscan Pro, Shining 3D) collects mesh data of the morphed origami structures. We write a MATLAB code to automatically collect individual panels' center positions from the scanned point cloud data, obtaining the global curvature of morphed origami structures.

**Numerical simulations:** Using ABAQUS Standard/CAE 6.14, we simulate the self-folding actuation by heating. We also use a quasi-static simulation of bistable properties of origami assemblies in Section 4 using the explicit module of ABAQUS. All parts are meshed with an S4R element in ABAQUS. See Supplementary Information for detailed simulation settings.

## Acknowledgments

This research was supported by the general program of the National Natural Science Foundation of China [Grant No. 12272225] and an International Collaboration Grant by the Ministry of Science and Technology in China [Grant No. SQ2022YFE010363]. The authors also acknowledge the financial support from National Natural Science Foundation of China (Projects 52375022, 52192631), and the International Collaboration Program of Tianjin University. The authors thank Ms. Bihui Zou and Mr. Zhipeng Liu for their technical discussions.

## Author contributions

J. J. and J. M. designed the research; Z. L. and S. C. designed the structures, performed the experiments, conducted the FE simulations, and analyzed the data; Z. L. designed the 4D printing technique and developed the theoretical model; Z. L., Q. D., and K. X. collected and processed the experimental data; K. L., S. C., and Z. L. analyzed the bistability qualitatively and quantitatively; and J. J. and J. M. provided guidance throughout the research. All authors participated in writing and reviewing the manuscript.